\documentclass[11pt,letterpaper]{article}

\usepackage{url,graphicx,amsmath}
\usepackage[american]{babel}
\usepackage[autostyle]{csquotes}
\usepackage[authordate,backend=biber,noibid,autolang=none,booklongxref=false,%
bibencoding=latin1,postnotepunct,compresspages,strict,%
annotation]{biblatex-chicago}
\usepackage{lmodern}
\usepackage{ifthen}
\usepackage{setspace}
\usepackage{graphicx}
\usepackage[usenames,dvipsnames]{color}
\usepackage{caption}
\usepackage{subcaption}
\usepackage[toc,page]{appendix}
\usepackage[letterpaper,nomarginpar,margin=1in]{geometry}
\usepackage[flushleft]{threeparttable}
\bibliography{bibHotGoalie}

\begin{document}

\title{Do NHL goalies get hot in the playoffs?\\A multilevel logistic regression analysis}
\author{Likang Ding, Ivor Cribben, Armann Ingolfsson, Monica Tran\\
University of Alberta School of Business \\
likang@ualberta.ca,~cribben@ualberta.ca,~armann.ingolfsson@ualberta.ca,~mt4@ualberta.ca}

\maketitle

\abstract{The hot-hand theory posits that an athlete who has performed well in the recent past performs better in the present. We use multilevel logistic regression to test this theory for National Hockey League playoff goaltenders, controlling for a variety of shot-related and game-related characteristics. Our data consists of $48,431$ shots for 93 goaltenders in the 2008--2016 playoffs. Using a wide range of shot-based windows to quantify recent save performance, we find no evidence for the hot-hand theory, and some evidence that good recent save performance negatively impacts the next-shot save probability. We use a permutation test to rule out a regression to the mean explanation for our findings.}


\section{Introduction}\label{sec:intro}

The hot-hand phenomenon generally refers to an athlete who has performed well in the recent past performing better in the present. Having a ``hot goalie'' is seen as crucial to success in the National Hockey League (NHL) playoffs. A goaltender who keeps all pucks out of the net for 16 games (4 series of 4 wins) will win his team the Stanley Cup---obviously. In this paper, we use data from the NHL playoffs to investigate whether goaltenders get hot, in the sense that if a goaltender has had a high recent save probability, then that goaltender will have a high save probability for the next shot that he faces.

NHL fans, coaches, and players appear to believe that goaltenders can get hot and that a hot goaltender can be the key to the success of an NHL team. A famous example of a hot goaltender is the legendary playoff journey of Jonathan Quick of the Los Angeles Kings in 2012. With the fifth-best save percentage ($92.86\%$) in the regular season, Quick continued to improve ($95.35\%$) in the first round of the playoffs against the Vancouver Canucks. He remained hot until lifting the first Stanley Cup in franchise history and completed the playoffs with the best save percentage ($94.61\%$) of all goalies who played more than 10 playoff games.

NHL goaltenders let in roughly one in ten shots. More precisely, during the 2018-19 regular season, 93 goaltenders playing for 31 teams faced a total of 79,540 shots, which resulted in 7,169 goals, for an average save percentage of 90.99\%. Among goaltenders who played at least 20 games, the season-long save percentage varied from a high of 93.42\% (Ben Bishop, Dallas Stars) to a low of 88.58\% (Aaron Dell, San Jose Sharks)—a range from 2.41 percentage points (pps) below to 2.43 pps above the overall average save percentage. In the playoffs of the same year, the overall average save percentage was 91.64\%. Among goaltenders who played in two or more playoff games, the save percentage varied from 93.56\% (Robin Lehner, New York Islanders) to 85.58\% (Andrei Vasilevskiy, Tampa Bay Lightning)—--a range from 6.06 pps below to 1.92 pps above the overall average save percentage.

It is crucial to determine whether the hot-hand phenomenon is real, for NHL goaltenders, to understand whether coaches are justified in making decisions about which of a team's two goaltenders should start a particular game based on perceptions or estimates of whether that goaltender is hot. If the hot hand is real, then appropriate statistical models could potentially be used to predict the likely performance of a team's two goaltenders in an upcoming game, or even in the remainder of a game that is in progress, during which the goaltender currently on the ice has performed poorly.  

Our main finding is the non-existence of the hot-hand phenomenon, that is, no positive influence of recent save performance on the save probability for the next shot. We also find a statistically significant \emph{negative} slope coefficient for the influence of recent save performance on the save probability for one window size; in other words, we demonstrate that contrary to the hot-hand theory, better past performance may result in a \emph{worse} future performance. For other window sizes, we find insignificant but negative slope coefficients. A permutation test indicates that our main finding is not simply the result of regression-to-the-mean bias.

The remainder of the paper is organized as follows: in Section \ref{sec:LR}, we review related literature; in Section \ref{sec:data}, we describe our data set; in Section \ref{sec:method}, we specify our regression models; in Sections \ref{sec:results} and \ref{sec: permutation}, we present our results; and in Section \ref{sec:conclusion}, we discuss our findings.

\section{Literature review}\label{sec:LR}

We summarize five streams of related work addressing the following: (1) whether the hot hand is a real phenomenon or a fallacy, (2) whether statistical methods have sufficient power to detect a hot hand, (3) whether offensive and defensive adjustments reduce the impact of a hot hand, (4) estimation of a hot-hand effect for different positions in a variety of sports, and (5) specification of statistical models to estimate the hot hand.

\emph{(1) Is the hot hand a real phenomenon or a fallacy?} The hot hand is originally studied in the 1980s in the context of basketball shooting percentages \autocite{gilovich1985hot,tversky1989hot,tversky1989cold}. These studies conclude that even though players, coaches, and fans all believe strongly in a hot-hand phenomenon, no convincing statistical evidence supports its existence. Instead, \textcite{gilovich1985hot} attribute beliefs in a hot hand to a psychological tendency to see patterns in random data; an explanation that has also been proposed for behavioral anomalies in various non-sports contexts, such as financial markets and gambling \autocite{miller2018surprised}.  Contrary to these findings, recent papers by \textcite{miller2018surprised,miller2019bridge} demonstrate that the statistical methods used in the original studies are biased, and when their data is re-analyzed after correcting for the bias, strong evidence for a hot hand emerges.

\emph{(2) Do statistical methods have sufficient power to detect a hot hand?} \textcite{gilovich1985hot} analyze players individually. This approach may lack sufficient statistical power to detect a hot hand, even if it exists \autocite{Wardrop1995,wardrop1999statistical}.  Multivariate approaches that pool data for multiple players have been proposed to increase power \autocite{arkes2010revisiting}. We follow this approach, by pooling playoff data for multiple NHL goaltenders over multiple years.

\emph{(3) Do offensive and defensive adjustments reduce the impact of a hot hand?} 
A hot hand, even if it is real, may not result in measurable improvement in performance if the hot player adapts by making riskier moves or if the opposing team transfers resources to focus on the player. For example, a hot basketball player may attempt riskier shots and the opposing team may double team such a player. The extent to which such adjustments can be made varies by sport, by position, and by the game situation: There is little opportunity for such adjustments for basketball free throws \autocite{gilovich1985hot} and there is less opportunity to transfer defensive resources towards a single player in baseball than in basketball \autocite{green2018hot} because the fielding team faces batters sequentially. Resources can only be transferred across time by letting better pitchers face better batters, rather than across space for the defensive team since there is only one batter in the batter's box at a time. The opportunity for transferring resources is even less for an NHL team facing a hot goalie than for a baseball team facing a hot batter, because the NHL team typically faces the same goalie for the entire game, which minimizes the opportunity to transfer resources across time. Transferring resources across space is only possible through strategies like ``crowding the net''. Overall, the opportunities to transfer resources away from other tasks and towards scoring are more limited than in the case of basketball and even baseball. NHL goaltenders thus provide an ideal setting in which to measure whether the hot-hand phenomenon occurs.

\emph{(4) Estimation of a hot-hand effect for different positions in a variety of sports.}  In addition to basketball shooters, the list of sports positions for which we have investigated hot-hand effects includes baseball batters and pitchers \autocite{green2018hot}, soccer penalty shooters \autocite{otting2022regularized}, soccer in-game scoring \autocite{parsons2015hot}, dart players \autocite{otting2020hot}, and golfers \autocite{livingston2012hot}. 

In ice hockey, a momentum effect has been investigated at the team level \autocite{kniffin2014within}.  \textcite{vesper2015putting} has investigated hot-hand effects for ice hockey shooters but not for goaltenders, except for the study by \textcite{morrison1998takes}. The latter study focuses on the duration of NHL playoff series, noted a higher-than-expected number of short series, and proposed a goaltender hot-hand effect as a possible explanation. This study has not analyzed shot-level data for goaltenders, as we do.

\emph{(5) Specification of statistical models to estimate the hot hand.}  Hot-hand researchers have used two main approaches in specifying their statistical models: (1) Analyze success rates, conditional on outcomes of previous attempts \autocite{albright1993statistical,green2018hot} or (2) incorporate a latent variable or ``state'' that quantifies ``hotness'' \autocite{green2018hot,otting2022regularized}. We follow the former approach. With that approach, past performance is typically summarized over a ``window'' defined in terms of a fixed number of past attempts---the ``window size.''  It is not clear how to choose the window size. We vary the window size over a range that covers the window sizes used in past work. We also perform the analysis using time-based windows, an approach that complicates data preparation and has not been used by other investigators. Results for the two windows types are consistent (Appendix \ref{appendix: time based window}).

We contribute to the hot-hand literature by investigating NHL goaltenders. This position has not been studied previously and provides a setting in which there are limited opportunities for either team to adapt their strategies in reaction to a perception that a goaltender is hot. In terms of methodology, we use multilevel logistic regression, which allows us to pool data across goaltender-seasons to increase statistical power, and we use a wide range of shot-based windows to quantify a goaltender's recent save performance.

\section{Data and variables}\label{sec:data}

Our data set consists of information about all shots on goal in the NHL playoffs from 2008 to 2016. The season-level data is from \url{www.hockey-reference.com} \autocite{HockeyReference} and the shot-level data is from \url{corsica.hockey} \autocite{Corsica}. We have data for 48,431 shots, faced by 93 goaltenders, over 795 games and 9 playoff seasons, with an average of 30.46 shots on goal per team per game and 91.64\% of the shots resulting in a save.  We divide the data into 224 groups, containing from 2 to 849 shot observations, based on combinations of goaltender and playoff season. The data set includes 1,662 shot observations for which one or more variables have missing values. Removing those observations changes the average save proportion from 91.64\% to 91.61\% and the number of groups from 224 to 223. We exclude observations with missing values from our regression analysis but we include these observations when computing the variable of interest (recent save performance), as discussed in Section \ref{Saving}.

\subsection{Dependent variable: Shot outcome}
The dependent variable, $y_{ij}$, equals 1 if shot $i$ in group $j$ results in a save and 0 if the shot resulted in a goal for the opposing team. Following NHL convention \autocite{NHL.com}, a shot that hits the crossbar or one of the goalposts, or is blocked by players other than the goalie, is not counted as a shot on goal and is not included in our data set.

\subsection{Variable of interest: Recent save performance}\label{Saving}

The primary independent variable of interest, $x_{ij}$, is the recent save performance, immediately before shot $i$ in group $j$. It is not obvious how to quantify this variable so we investigate several possibilities. In all cases, we define the recent save performance as the ratio of the number of saves to the number of shots faced by the goaltender, over some ``window''. The window could be defined in terms of the number of shots faced, the amount of time played, or the amount of time elapsed. Time-based windows have the disadvantage that the variability of $x_{ij}$ depends on the number of shots in the window. Therefore, we focus on shot-based windows. We use a set of window sizes $K = \{30, 60,90, 120, 150\}$ shots, which correspond, roughly, to $\{1, 2, 3, 4, 5\}$ games.  Our longest window corresponds to the 5-game interval that \textcite{green2018hot} suggest is needed to determine whether a baseball player is hot.

For each group $j$, we set the shot index $i$ equal to 0 for the last shot in the regular season and 1 for the first shot in the playoffs. Thus, $i \geq 1$ for shots during the playoffs and $i \leq 0$ for shots during the regular season. For a shot-based window of length $k \in K$, we compute $x_{ij}$ as:
\begin{align}
    x_{ij}& = \frac{1}{k} \sum_{n = i - k}^{i -1}y_{nj}.\label{eqn: definition shot based window x_ij}
\end{align}

A window could include one or more intervals during which the group $j$ goaltender was replaced by a backup goaltender. Shots faced by the backup goaltender are excluded from the computation of $x_{ij}$. A window could include time periods between consecutive games, which could last several days.

As stated previously, we included shots with missing values for the control variables in the computation of $x_{ij}$ but we excluded those shots from the regression analysis that we describe in Section \ref{sec:method}. We also describe other excluded shots in Section \ref{sec:method}.

\subsection{Control variables: Other influential factors}
We include a vector, $Z_{ij}$, of eleven control variables for shot $i$ of group $j$ that we expect to impact a shot outcome. The control variables are: {\sf Angle}, {\sf Distance}, {\sf Game score}, {\sf Home}, {\sf Period}, {\sf Rebound}, {\sf $>$ 60 shots in the last 4 days}, {\sf Strength}, {\sf Shot type}, {\sf Opponent points percentage}, and {\sf Round}. {\sf Angle}, {\sf Distance}, and {\sf Opponent points percentage} are numerical variables and the rest are categorical. In what follows, we elaborate on each of the control variables.

\begin{figure}
	\centering\caption{{\sf Angle} $= \alpha$ and {\sf Distance} $=d$ for shot origin.}\label{CST}
    \includegraphics[width=4cm]{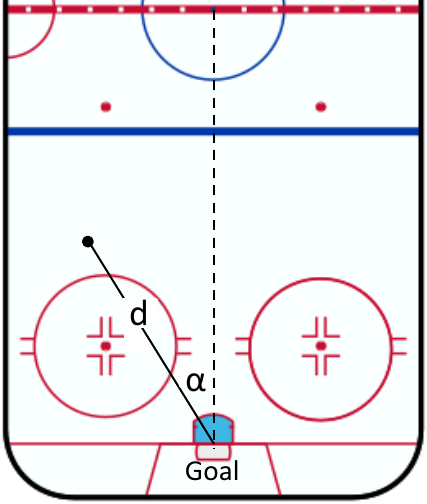}
\end{figure}

{\sf Angle} (in degrees, from 0 to 90) and {\sf Distance} (in feet) are measured based on a line from the shot origin to the midpoint of the crossbar of the goal (see Figure \ref{CST}). We only have the absolute value of the {\sf Angle}, that is, the data does not indicate whether the shot is from the left or the right wing. Preliminary analysis indicates that the log odds of the save probability is, approximately, a linear function of {\sf Angle} and {\sf Distance}.

{\sf Game score} indicates whether the goaltender's team is Leading (base case), Tied, or Trailing in the game. {\sf Home} is a binary variable indicating whether the goaltender is playing on home ice or not (base case). We code the {\sf Period} of the game as 1 (base case), 2, 3, Overtime. {\sf Rebound} is a binary variable indicating whether the shot occurs within 2 seconds of uninterrupted game time of another shot from the same team \autocite{Corsica} or not (base case). The binary {\sf $>$ 60 shots in the last 4 days} variable (base case: $\leq 60$) is intended to capture goalie fatigue. This roughly corresponds to playing two full consecutive games within the preceding four calendar days.  {\sf Strength} represents the difference between the number of players from the goaltender's team on the ice and the number of players from the other team on the ice, and takes values of $+$, $0$ (base case), or $-$. {\sf Shot type} denotes the shot type: Backhand (base case), Deflected, Slap, Snap, Tip-in, Wrap-around, or Wrist. {\sf Opponent points percentage} is the opponent team's regular season percentage of points earned, and we use it to control for the quality of the opponent team. {\sf Round} is the playoff round: 1 (base case), 2, 3, or 4, which correspond to the first rounds, semi-conference finals, conference finals, and finals. We note that the values of {\sf Home}, {\sf Opponent points percentage}, and {\sf Round} remain constant during the game to which shot $i$ of group $j$ belongs.

\section{Models}\label{sec:method}

We use multilevel logistic regression with partial pooling, also referred to as mixed effects modeling. We rescale the variable of interest $x_{ij}$ in (\ref{eqn: definition shot based window x_ij}) \autocite{enders2007centering, hox2011handbook} by subtracting the long-term average performance $\Tilde{x}_{ij}$ of the goaltender to obtain deviations from his performance: 
\begin{align}
    \Tilde{x}_{ij}& = \frac{1}{840} \sum_{n = i - 840}^{i -1}y_{nj}\label{eqn: long term shot based window x_ij}.
\end{align}

We compute $\Tilde{x}_{ij}$ as in (\ref{eqn: definition shot based window x_ij}), but with a longer window size: $k = 28 \times 30 = 840$. This window size corresponds, roughly, to 28 games, which is the longest possible playoff journey (7 games $\times$ 4 rounds). It is not straightforward to determine whether a positive deviation in player performance should be attributed to ``being hot'' versus ``being a much improved player'' \autocite{green2018hot}. If a player's performance increases over time, then a long-term performance estimate based on a long time window (say, the career average) will lag behind the player's current ability, resulting in a positive deviation that should not be attributed to being hot. To avoid this issue, we use a window size that is longer than the largest of the window sizes for calculating recent performance, but short enough so that the player's ability is unlikely to change much.

We exclude all shots from the regression model for which $\tilde{x}_{ij}$ cannot be computed because the goaltender did not play enough during the current season to compute the long-term average performance. Specifically, (1) for groups in which the goaltender's regular season play exceeds the long-term window size, we include all shots; (2) for groups in which the goaltender's regular season and playoff play combined fall short of the long-term window size, we exclude all shots; and (3) for groups that fall between these two extremes, we exclude some shots. In total, we exclude 7,166 shots ($14.80\%$) for the shot-based window analysis. 

Our variable of interest becomes:
\begin{align}
    x^\prime_{ij}&= x_{ij} - \Tilde{x}_{ij}. \label{eqn: rescaling of x_ij}
\end{align}
Our interest is in whether deviations from long-term performance persist over time.

We allow the intercept and the slope coefficient of the variable of interest to vary by group, but the control variable slope coefficients are the same for all groups, as shown in the following partial pooling specification:
\begin{align}
	\Pr(y_{ij}=1) = \mbox{logit}^{-1} (\alpha_{j} + \beta_{j}x_{ij}^\prime + \gamma Z_{ij}),
\end{align}
\noindent where $\mbox{logit}(p) \equiv \ln(p/(1-p))$, $\alpha_{j}$ is the intercept for group $j$, $\beta_{j}$ is the slope for group $j$, and $\gamma$ is the global vector of coefficients for the control variables. We represent the intercept and slope of the variable of interest as the sum of a fixed effect and a random effect, that is: $\alpha_j = \bar{\alpha} + \alpha_j^\ast$ and $\beta_j = \bar{\beta} + \beta_j^\ast$. 

All estimates that we report in Section \ref{sec:results} are obtained using Markov chain Monte Carlo (MCMC), using the {\sf rstan} and {\sf rstanarm} R packages. We use the default prior distributions from the {\sf rstanarm} package. The default distributions are weakly informative---Normal distributions with mean $0$ and scale $2.5$ (for the slope coefficients) and $10$ (for the intercepts) and multivariate Normal distributions with zero mean vector and an unknown covariance matrix to be estimated for the random effects \autocite{PriorDistributions}. We use default values for the number of chains ($4$), the number of iterations (1,000), and the burn-in (1,000 iterations).

\section{Results}\label{sec:results}
First, we provide detailed results for the $k = 120$ window. Second, we investigate the consistency of our main finding by varying the window size. Third, we illustrate the effect size of all independent variables (the variable of interest and all control variables) on the save probability. Appendix \ref{appendix: time based window} provides additional results for time-based windows; Appendix \ref{appendix: MCMC diagnostics} provides diagnostics for the MCMC estimation.

\subsection{Baseline results}
Table \ref{table:Results_3} provides means and 90\% credible intervals (Bayesian confidence intervals) for the intercept and slope fixed effects ($\bar{\alpha}$ and $\bar{\beta}$) and for the control variable coefficients ($\boldsymbol{\gamma}$), for our baseline model: the model with 120-shot window size.

\begin{table}
\centering
\begin{threeparttable}
\caption{Results for the $k = 120$ model.}\label{table:Results_3}
\begin{tabular}{lr@{}lrr}
 &\multicolumn{4}{c}{\textbf{$k = 120$ shots}} \\

	Variable 
	& \multicolumn{2}{r}{Mean} & 5\% & 95\% \\
	\hline

	Intercept
	&1&.08&0.62&1.54
 \\
	\textbf{Recent save performance ($x_{ij}^\prime$)}
	&$\mathbf{-1}$&$\mathbf{.82^\ast}$&-3.46&-0.18
	\\
 
	{\sf Angle} 
	&$0$&$.02^\ast$&0.01&0.02
	\\
 
	{\sf Distance} 
	&$0$&$.05^\ast$&0.05&0.06
	\\
	{\sf Game score: Tied} 
	&$-0$&$.05$&-0.13&0.03
	\\
	{\sf Game score: Trailing} 
	&$-0$&$.02$&-0.10&0.06
	\\
	{\sf Home} 
	&$-0$&$.07^\ast$&-0.13&-0.01
 	\\
        {\sf Opponent points percentage} 
	&$0$&$.15$&-0.57&0.83
        \\
        {\sf Period: 2}
        &$-0$&$.08^\ast$&-0.16&-0.01
        \\
        {\sf Period: 3}
       &$-0$&$.11^\ast$&-0.19&-0.02
        \\
        {\sf Period: Overtime}
        &$-0$&$.07$&-0.21&0.07
        \\
	{\sf Rebound} 
	&$-0$&$.74^\ast$&-0.84&-0.63
	\\
        {\sf $>60$ shots in the last $4$ days}
        &$0$&$.03$&-0.08&0.14
        \\
        {\sf Round: $2$} 
	&$-0$&$.06$&-0.14&0.02
        \\
        {\sf Round: $3$} 
	&$-0$&$.17^\ast$&-0.28&-0.06
        \\
        {\sf Round: $4$} 
	&$-0$&$.07$&-0.21&0.07
	\\
	{\sf Shot type: Deflected} 
	&$-0$&$.82^\ast$&-1.04&-0.60
	\\
	{\sf Shot type: Slap} 
	&$-0$&$.90^\ast$&-1.05&-0.77
	\\
	{\sf Shot type: Snap} 
	&$-0$&$.75^\ast$&-0.88&-0.61
	\\
	{\sf Shot type: Tip-in} 
	&$-0$&$.42^\ast$&-0.56&-0.29
	\\
	{\sf Shot type: Wrap-around} 
	&$0$&$.58^\ast$&0.26&0.89
	\\
	{\sf Shot type: Wrist} 
	&$-0$&$.39^\ast$&-0.50&-0.28
        \\
	{\sf Strength: $+$} 
	&$-0$&$.04$&-0.22&0.16
	\\
	{\sf Strength: $-$} 
	&$-0$&$.44^\ast$&-0.51&-0.37
	\\
\end{tabular}
 \begin{tablenotes}
       \item [$\ast$] $90\%$ credible interval excludes zero.
    \end{tablenotes}
\end{threeparttable}	
\end{table}

Our main finding from the baseline model is that a goaltender's recent save performance does not have a positive impact on the save probability, which shows the non-existence of the hot-hand effect; moreover, we report a \textit{negative} and statistically significant fixed effect that is contrary to the hot-hand theory.

Most of the control variables have a significant impact on the save probability. The posterior mean values for the significant control variables are in the direction we expect, except for Period 3 and {\sf Round}. Specifically, the posterior means indicate that, compared to baseline values, a goaltender performs better if his team is not on a penalty kill, on home ice, facing a wrap-around shot, and facing a shot from a long distance or large angle. By contrast, a goaltender performs worse in Periods 2 and 3, in Round 3, facing any shot type other than backhand and wrap-around, and facing a rebound shot. The poor performance in Period 3 is unexpected. The expected sign for {\sf Round} is unclear.

\subsection{Consistency of the main finding}
\begin{figure}
\begin{minipage}[t]{0.45\linewidth}
\includegraphics[width=.8\linewidth]{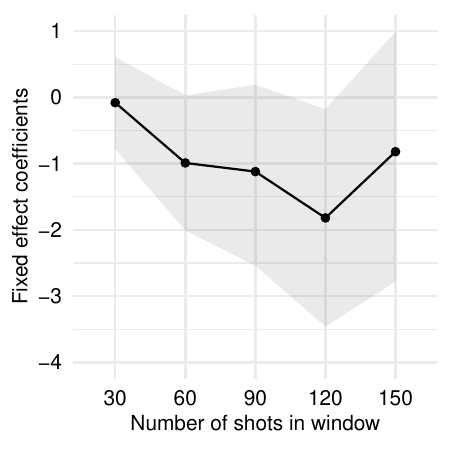}
\caption{Recent save performance fixed effect coefficients ($\hat{\bar{\beta}}$s) with $90\%$ credible intervals for all window sizes.}\label{fig: fixed effect estimates}
\end{minipage}
\hfill
\begin{minipage}[t]{0.45\linewidth}
\includegraphics[width=.8\linewidth]{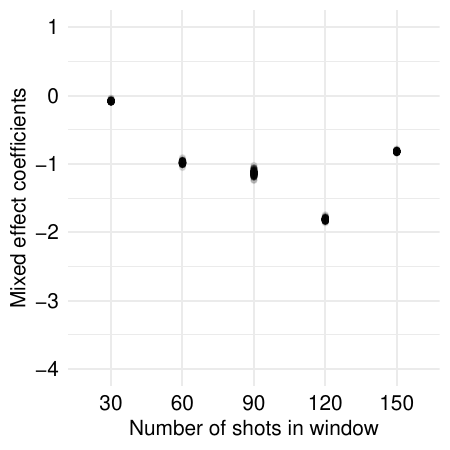}
\caption{Distributions for recent save performance mixed effects ($\hat{\beta}_j$s) for all window sizes.}\label{fig: group effect comparison}
\end{minipage}%
\end{figure}

Our main finding, that recent save performance has a negative fixed effect value, holds for all window sizes (Figure \ref{fig: fixed effect estimates}). However, the fixed effect is statistically significant at the $10\%$ level only for the $k = 120$ window (see $90\%$ credible intervals in Figure \ref{fig: fixed effect estimates}).

The fact that the slope fixed effects for the recent save performance, $\hat{\bar{\beta}}$, are negative leaves open the possibility that the slopes for some individual goaltender-seasons are positive. But Figure \ref{fig: group effect comparison} shows that all of the individual-group slope coefficients, $\hat{\beta}_j$ are negative and closely clustered around the fixed effects, for all window sizes.

\begin{figure}
	\centering
	\caption{Control variable slope coefficients ($\hat{\gamma}$s), for all window sizes. The point estimates are represented by blue crosses. $90\%$ credible intervals from the 120-shot baseline model are included for comparison.}\label{ControlGraph}
	\includegraphics[width=10cm]{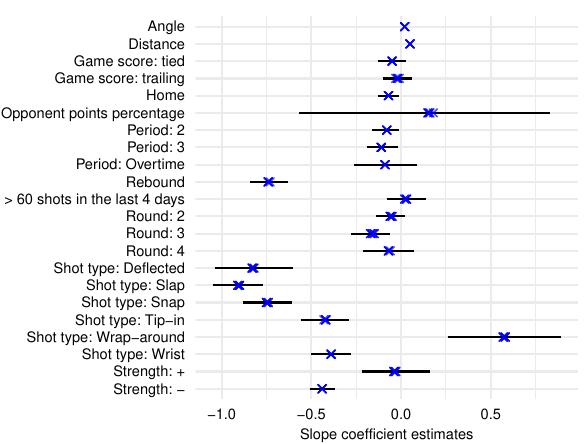}
\end{figure}
The slope coefficients for all control variables are consistent in sign and magnitude and are within the 120-shot model $90\%$ credible interval, for all window sizes (Figure \ref{ControlGraph}). Note that the credible intervals for {\sf Angle} and {\sf Distance} are so narrow that they are not visible.

\subsection{Effect size of independent variables on save probability}
We investigate the impact of recent save performance and the control variables on the estimated save probability for the next shot, using the baseline model. To calculate the impact of varying a particular variable on the save probability, we set all other independent variables to their average values (sample proportions for categorical variables). We consider the average value for $x^\prime_{ij}$ to be 0.

The impact of increasing the recent save performance from $4.76\%$ below to $3.93\%$ above the long-term performance (corresponds to moving from the 2.5th to the 97.5th percentile) is a 0.91 pp decrease in the estimated save probability for the next shot. For comparison, this number is about one-fifth of the 4.84 pp range of average save percentages during the 2018-19 regular season (as discussed in Section \ref{sec:intro}).

Given that we define $x_{ij}^\prime$ to be the \emph{deviation} in performance of the goaltender-season group $j$ from his long-term performance, the percentiles for $x_{ij}^\prime$ correspond to different recent save performances for different groups. To illustrate the effect more concretely, consider the largest group: the 832 shots faced by Tim Thomas during the 2011 playoffs. His long-term performance was $92.98\%$ (as of the end of the regular season). We replace the values of control variables with their group means in calculating save probabilities. For shots where Thomas' recent save performance deviation was at the 2.5th percentile, corresponding to $x_{ij} = 90.83$\%, his estimated save probability for the next shot was $95.76\%$---a $4.93$ pp increase. At the other extreme, for shots where the recent save performance was at the 97.5th percentile, corresponding to $x_{ij} = 97.50$\%, the estimated save probability for the next shot was 95.29\%---a 2.21 pp decrease.

\begin{figure}
 \caption{Control Variables versus estimated save probability}\label{ControlProb}
 \centering
 \begin{subfigure}[b]{0.5\textwidth}
 \includegraphics[width=\linewidth]{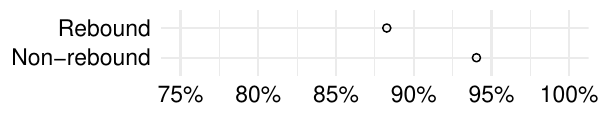}
 \caption{{\sf Rebound}}
 \end{subfigure}%
 \begin{subfigure}[b]{0.5\textwidth}
 \includegraphics[width=\linewidth]{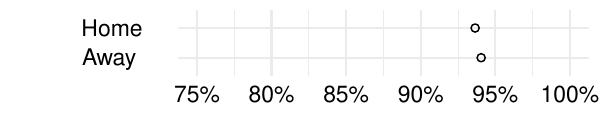}
 \caption{{\sf Home}}
 \end{subfigure}%
 
 \begin{subfigure}[b]{0.5\textwidth}
 \includegraphics[width=\linewidth]{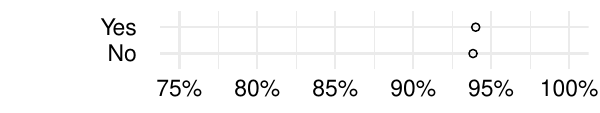}
 \caption{{\sf $>60$ shots in the last $4$ days}}
 \end{subfigure}%
 \begin{subfigure}[b]{0.5\textwidth}
 \includegraphics[width=\linewidth]{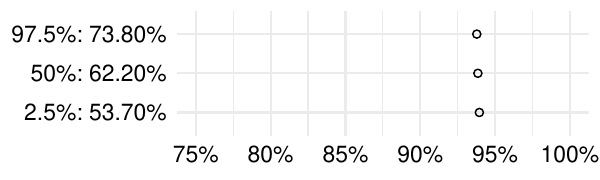}
 \caption{{\sf Opponent points percentage}}
 \end{subfigure}%
 
 \begin{subfigure}[b]{0.5\textwidth}
 \includegraphics[width=\linewidth]{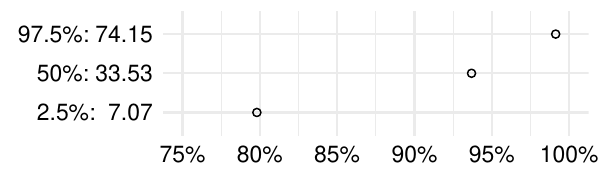}
 \caption{{\sf Distance (in feet)}}
 \end{subfigure}%
 \begin{subfigure}[b]{0.5\textwidth}
 \includegraphics[width=\linewidth]{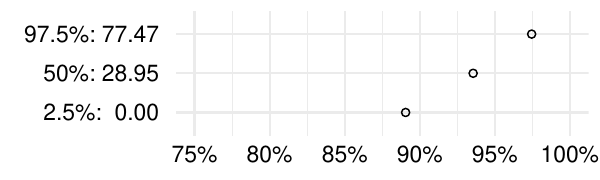}
 \caption{{\sf Angle (in degrees)}}
 \end{subfigure}%
 
 \begin{subfigure}[b]{0.5\textwidth}
 \includegraphics[width=\linewidth]{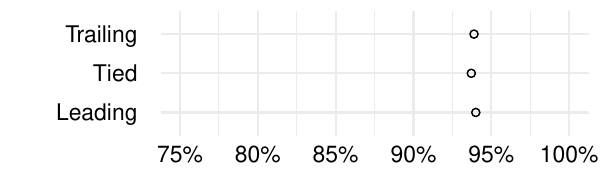}
 \caption{{\sf Game score}}
 \end{subfigure}%
 \begin{subfigure}[b]{0.5\textwidth}
 \includegraphics[width=\linewidth]{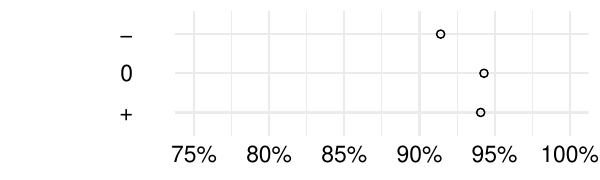}
 \caption{{\sf Strength}}
 \end{subfigure}%
 
 \begin{subfigure}[b]{0.5\textwidth}
 \includegraphics[width=\linewidth]{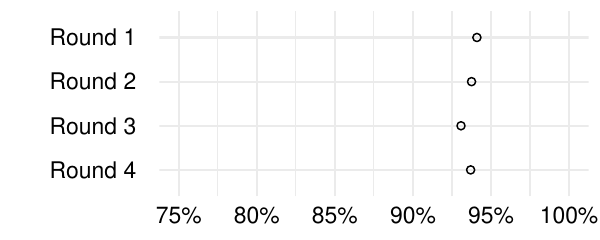}
 \caption{{\sf Round}}
 \end{subfigure}%
 \begin{subfigure}[b]{0.5\textwidth}
 \includegraphics[width=\linewidth]{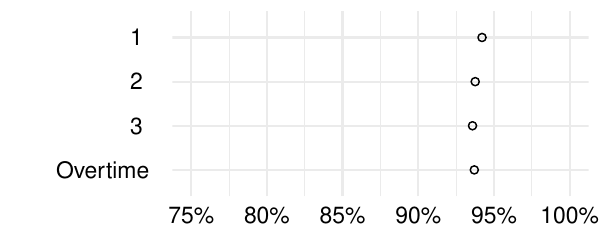}
 \caption{{\sf Period}}
 \end{subfigure}%
 
 \begin{subfigure}[b]{0.5\textwidth}
 \includegraphics[width=\linewidth]{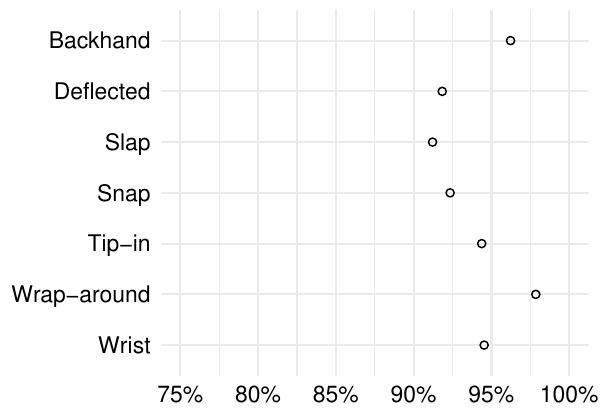} 
 \caption{{\sf Shot type}}
 \end{subfigure}%

\end{figure}

Figure \ref{ControlProb} shows the impact of the control variables. {\sf Rebound}, {\sf Distance}, {\sf Angle}, {\sf Strength}, and {\sf Shot type} have a substantial impact on estimated save probability: a rebound shot is 6 pps less likely to be saved than a non-rebound shot; a shot from 74 feet is 20 pps less likely to result in a goal than a shot from 7 feet; a shot from 77 degrees is 8 pps more likely to be saved than a shot from 0 degrees; and a shot faced by a team on the power play is 3 pps more likely to be saved compared with a team on penalty kill. For {\sf Shot type}, the save probability decreases by 8 pps when moving from wrap-around (the shot type least likely to result in a goal) to a slap shot (the type most likely to result in a goal). The other control variables have a minor impact.

\section{Permutation test for estimation bias}\label{sec: permutation}
In our rescaling process, we subtract a long-term average over a window that only includes shots that occur {\it before} shot $i$. This approach, however, may introduce a regression-to-the-mean effect which causes negative fixed effect estimates: if a goaltender's performance, after controlling for observable factors, is completely random, then we will expect a period of above-average or below-average recent save performance to be likely to be followed by a period of save performance that is closer to the average. To determine whether the negative $\hat{\bar{\beta}}$ is only from the regression-to-the-mean effect, we present a permutation test to determine whether this rescaling process would introduce significant negative biases.

For this analysis, we use the 120-shot baseline model and perform 100 permutations. In each permutation, we randomly permute the $y_{ij}$ values across the goalie-season groups, and we recalculate the $x_{ij}$ and $\Tilde{x}_{ij}$ values using the permuted $y_{ij}$ values. By randomly permuting the $y_{ij}$ values, we eliminate any correlations among the shot outcomes. If rescaling $x_{ij}$ values does not introduce any biases, then we expect the empirical distribution of estimated fixed effects to be centered around 0.

\begin{figure}
    \centering
    \caption{Results of permutation test}
    \includegraphics[width = 10.5 cm]{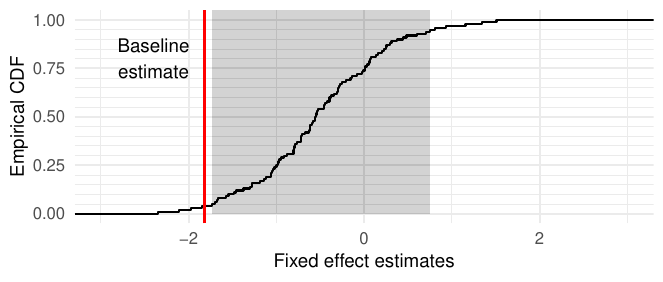}
    \label{fig: permutation test results}
\end{figure}

Figure \ref{fig: permutation test results} shows the results. The mean and median $\hat{\bar{\beta}}$ values are $-0.51$ and $-0.55$, suggesting a small negative bias. However, the recent save performance coefficient for the un-permuted data, of $-1.82$, is outside a $90\%$ interval defined by the 5th and 95th percentiles of the permutation distribution. We conclude that the negative fixed effect estimate that we obtained cannot be explained solely by a negative bias from the re-scaling process.

\section{Discussion and conclusion}\label{sec:conclusion}

We employ multilevel logistic regression to investigate whether the performance of NHL goaltenders during the playoffs is consistent with a hot-hand effect, using data from the 2008--2016 NHL playoffs. We measure past performance using shot-based windows. Our window sizes span a wide range: from, roughly, 1 game to 5 games. We allow the intercept and the slope for recent save performance to vary across goaltender-season combinations.

We find no evidence of hot-hand effects and a significant \emph{negative} impact of recent save performance on the next-shot save probability, for one window size. The estimates were negative for all window sizes and goaltender-season groups. Our findings are contrary to the findings for baseball in \textcite{green2018hot}, who use a window size similar to our longest window, and hypothesize that hot hand effects should generally be observed for skilled activity.

Referring back to Section \ref{sec: permutation}, if the effect we find were entirely due to regression to the mean, then we would expect that as we increase the window size used to measure recent save performance, the average amount by which performance moves toward the average to decrease. The fact that we observe an increase in magnitude followed by a decrease (see Figure \ref{fig: fixed effect estimates}), together with our permutation test results, argues against our finding being driven by regression to the mean.

A motivation effect provides one possible explanation for our findings. That is, if a goaltender's recent save performance has been below his average for the current playoffs, then his motivation increases, resulting in increased effort and focus, causing the next-shot save probability to be higher. Conversely, if the recent save performance has been above average, then the goaltender's motivation, effort, and focus decrease, leading to a lower next-shot save probability. \textcite{belanger2013driven} find support for the first of these effects (greater performance after failure) for ``obsessively passionate individuals'' but have not found support for the second effect (worse performance after success) for such individuals. The study has found support for neither effect for ``harmoniously passionate individuals.'' These findings are consistent with Hall-of-Fame goaltender Ken Dryden's (\citeyear{dryden2019life}) sentiment that ``if a shot beats you, make sure you stop the next one, even if it is harder to stop than the one before.''  The psychological mechanisms underlying our findings warrant further study.

Although the estimated recent save performance coefficients are consistently negative, their magnitude varies. In particular, the magnitude increases sharply with the window size up to $k = 120$. We expect to see more reliable estimates with longer window sizes, but the increase in magnitude is surprising, given that we define the recent save performance as a scale-free save percentage.

A literal interpretation of our negative finding suggests a strategy that prefers a goalie with below-average recent save probability because his save probability for the next few shots is expected to improve. One should be cautious about using such a strategy, keeping in mind Goodhart's Law, that ``any observed statistical regularity will tend to collapse once pressure is placed upon it for control purposes'' \autocite{Chapter8GoodhartsLawitsoriginsmeaningandimplicationsformonetarypolicy}. For one thing, such a strategy would provide an incentive to underperform in non-essential games to earn more playing time in big games.

One limitation of our study is that, in defining windows, we ignore the time that passes between games. Past research, such as \textcite{green2018hot}, shares this limitation. This limitation can be particularly serious for backup goaltenders, for whom the interval between two successive appearances may be several days long.

\singlespacing
\printbibliography
\newpage
\begin{appendices}
\section{Time-based windows}\label{appendix: time based window}
In addition to shot-based windows, we investigate time-based windows in our analysis. Time-based window models differ in how $x_{ij}$ is defined and how it is re-scaled. Overall, we find consistent results for time-based window models. 

In time-based windows, similar to shot-based windows, $x_{ij}$ is the ratio of saved shots during a window before the current shot. We use a set of window size $T = \{ 30, 60, 120, 180, 240, 300\}$ minutes consisting of window lengths that are comparable to shot-based windows. 

For each group $j$, let $s(i)$ be the time at which shot $i$ occurred, with time set to zero at the beginning of the playoffs. Thus, time is positive during the playoffs and negative during the regular season. Define $r(s(i),t)$ to be the number of shots in the time window $[s(i) - t, s(i))$ (excluding shot $i$). We compute $x_{ij}$ as follows, for a time-based window of length $t \in T$:
\begin{align}
    x_{ij}& = \frac{1}{r(s(i),t)} \sum_{n = i - r(s(i),t)}^{i -1}y_{nj},\label{eqn: definition time based window x_ij}
\end{align}
\noindent and long-term performance is calculated using $t = 1680$ in (\ref{eqn: definition time based window x_ij}): 
\begin{align}
    \Tilde{x}_{ij}& = \frac{1}{r(s(i),1680)} \sum_{n = i - r(s(i),1680)}^{i -1}y_{nj}\label{eqn: long term time based window x_ij}.
\end{align}
Using the same criteria as in shot-based window analysis to exclude shots from the dataset, we exclude 6,664 ($13.76\%$) shots for the time-based window analysis. The rest of the model details are the same as for the shot-based window models.

\begin{figure}
\begin{minipage}[t]{0.45\linewidth}
 \includegraphics[width=.8\linewidth]{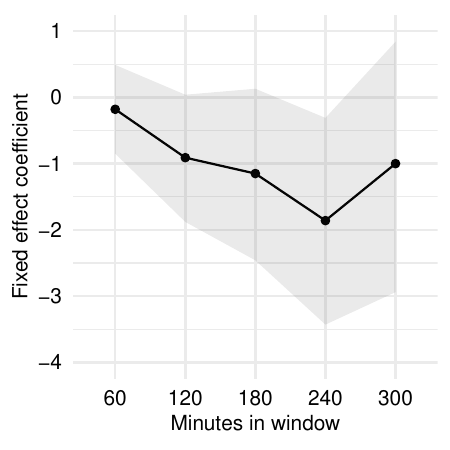}
\caption{Recent save performance fixed effect coefficients ($\hat{\bar{\beta}}$s) with $90\%$ credible intervals for all time-based window.}\label{fig: fixed effect estimates time based}
\end{minipage}
\hfill
\begin{minipage}[t]{0.45\linewidth}
\includegraphics[width=.8\linewidth]{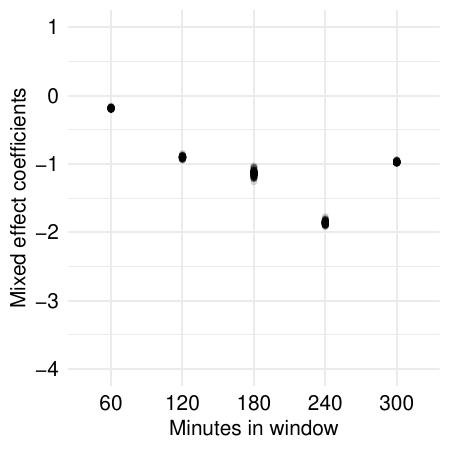}
\caption{Distributions for recent save performance mixed effects ($\hat{\beta}_j$s) for all time-based windows.}\label{fig: group effect comparison time based}
\end{minipage}%
\end{figure}

Results for time-based windows are similar to shot-based windows as illustrated in Figures \ref{fig: fixed effect estimates time based} and \ref{fig: group effect comparison time based}. Similar to shot-based windows, the fixed effect coefficients are all negative, and $\hat{\bar{\beta}}$ is only significant for the $240$-minute window, which is comparable to the 120-shot window; all of the individual-group slope coefficients  $\hat{\beta}_j$ are negative and clustered around the fixed effects as well.

\section{MCMC diagnostics}\label{appendix: MCMC diagnostics}
We computed two diagnostic statistics: $\hat{R}$ and $n_{\rm{eff}}$. To check whether a chain has converged to the equilibrium distribution, we can compare the chain’s behavior to other randomly initialized chains. The potential scale reduction statistic, $\hat{R}$, allows us to perform this comparison, by computing the ratio of the average variance of draws within each chain to the variance of the pooled draws across chains; if all chains are at equilibrium, the two variances are equal and $\hat{R} = 1$, and this is what we found for all of our models.

The effective sample size, $n_{\rm{eff}}$, is an estimate of the number of independent draws from the posterior distribution for the estimate of interest. The $n_{\rm{eff}}$ metric computed by the {\sf rstan} package is based on the ability of the draws to estimate the true mean value of the parameter. As the draws from a Markov chain are dependent, $n_{\rm{eff}}$ is usually smaller than the total number of draws. \textcite{gelman2013bayesian} recommend running the simulation until $n_{\rm eff}$ is at least 5 times the number of chains, or $5 \times 4 = 20$. This requirement is met for all of our models.
\end{appendices}
\end{document}